\documentstyle[12pt]{article}
\begin{document}\thispagestyle{empty}\begin{flushright}
OUT--4102--65\\hep-th/9612012\\
30 November 1996         \end{flushright}\vspace*{2mm}\begin{center}{\Large\bf
Conjectured Enumeration of irreducible Multiple  \\[5pt]
Zeta Values, from Knots and Feynman Diagrams     }\vglue 10mm{\large{\bf
D.~J.~Broadhurst$                                ^{1)}$}\vglue 4mm
Physics Department, Open University              \\[3pt]
Milton Keynes MK7 6AA, UK     }\end{center}\vfill\noindent{\bf Abstract}\quad
Multiple zeta values (MZVs) are under intense investigation in three arenas
-- knot theory, number theory, and quantum field theory -- which unite in
Kreimer's proposal that field theory assigns MZVs to positive knots, via
Feynman diagrams whose momentum flow is encoded by link diagrams. Two
challenging problems are posed by this nexus of knot/number/field theory:
enumeration of positive knots, and enumeration of irreducible MZVs. Both were
recently tackled by Broadhurst and Kreimer (BK). Here we report large-scale
analytical and numerical computations that test, with considerable severity,
the BK conjecture that the number, $D_{n,k}$, of irreducible MZVs of weight
$n$ and depth $k$, is generated by $\prod_{n\ge3}\prod_{k\ge1}(1-x^n y^k)
^{D_{n,k}}=1-\frac{x^3y}{1-x^2}+\frac{x^{12}y^2(1-y^2)}{(1-x^4)(1-x^6)}$,
which is here shown to be consistent with all shuffle identities for the
corresponding iterated integrals, up to weights $n=44,\,37,\,42,\,27$, at
depths $k=2,\,3,\,4,\,5$, respectively, entailing computation at the
petashuffle level. We recount the field-theoretic discoveries of MZVs,
in counterterms, and of Euler sums, from more general Feynman diagrams,
that led to this success.                     \vfill\footnoterule\noindent
$^1$) email: D.Broadhurst@open.ac.uk
\newpage\setcounter{page}{1}

\subsection*{1. Introduction}

In a recent Physics Letter~\cite{BK15} (hereafter BK), Dirk Kreimer and the
author made a guess, informed by field theory, at the number, $D_{n,k}$,
of irreducible multiple zeta values (MZVs)~\cite{DZ}
of weight $n$ and depth $k$ that enter a
minimal ${\bf Q}$-basis, to which all other MZVs may be reduced,
as rational combinations of elements of the basis set, and their products.
MZVs, and their extension~\cite{EUL} to alternating Euler sums,
concerned BK, as practitioners of perturbative quantum field theory (pQFT),
because of Kreimer's connection~\cite{DK1,DK2,4TR,DK} of MZVs with
positive knots~\cite{VJ}, via the counterterms of pQFT. This
connection is strongly
supported by BK's joint~\cite{BK15,BKP,BDK,BGK,BK4} and
separate~\cite{5LB,1440,Z2S6,DK3,IMU3} calculations of
multi-loop Feynman diagrams, and by those of
others workers~\cite{BKa,DKT,JJN,DEM}.

At first parallel~\cite{CK,LM,BBG} to, and recently
intertwining~\cite{BG,BBB,MEH,BBBR} with, this flurry of field-theoretic
activity, is an equally dramatic increase of understanding of MZVs~\cite{DZ}
and of alternating~\cite{EUL} Euler sums, as richly structured
number-theoretic objects, in their own right, irrespective of any
field or knot theory.
This is a rapidly progressing subject, where little is yet
rigorously proven, but much is conjectured, on the basis of
extensive~\cite{EUL,BBB} analytical and numerical computations.
Several conjectures command wide~\cite{DZ,EUL,BBG,BG,BBB,MEH} support.
Yet the answer to a very obvious question  --
how many MZVs are left undetermined by the
relations between MZVs? -- appeared to be difficult, even to guess.

As in~\cite{BK15}, we approach the question in three stages.
First, we consider Euler sums,
of weight $\sum_j s_j$ and depth $k$,
which are $k$-fold nested sums of the form~\cite{BK15,BBB}
\[\zeta(s_1,\ldots,s_k;\sigma_1,\ldots,\sigma_k)=\sum_{n_j>n_{j+1}>0}
\quad\prod_{j=1}^{k}\frac{\sigma_j^{n_j}}{n_j^{s_j}}\,,\]
with signs $\sigma_j=\pm1$, positive integer exponents $s_j$,
and $\sigma_1s_1\neq1$, to prevent divergence of the outermost sum.
As in~\cite{BBB}, we combine the strings of exponents and
signs into a single depth-length argument string,
with $s_j$ in the $j$th position when $\sigma_j=+1$, and $\overline{s}_j$
in the $j$th position when $\sigma_j=-1$.
The first question is then:\\[5pt]\indent{\bf Q1}\quad
What is the number, $E_{n,k}$, of Euler sums of weight $n$ and depth $k$,
in a\\\indent\phantom{{\bf Q1}\quad}
minimal {\bf Q}-basis for reducing all Euler sums to basic Euler sums?\\[5pt]
Next, we define MZVs~\cite{DZ} as non-alternating Euler sums,
with $\sigma_j=1$, and ask:
\\[5pt]\indent{\bf Q2}\quad
What is the number, $M_{n,k}$, of Euler sums of weight $n$ and depth $k$,
in a\\\indent\phantom{{\bf Q2}\quad}
minimal {\bf Q}-basis for reducing all MZVs to basic Euler sums?\\[5pt]
Finally comes the most natural, yet most difficult question:
\\[5pt]\indent{\bf Q3}\quad
What is the number, $D_{n,k}$, of MZVs of weight $n$ and depth $k$,
in a\\\indent\phantom{{\bf Q3}\quad}
minimal {\bf Q}-basis for reducing all MZVs to basic MZVs?

Section~2 gives the conjectured answers and focuses attention
on the most difficult question, {\bf Q3}, whose answer
followed discoveries~\cite{BK15,EUL,BGK} of how Feynman diagrams
assign MZVs to knots.
Section~3 characterizes the two types of identity that are believed
to achieve the conjectured reductions: depth-length shuffles,
and weight-length shuffles.
Section~4 presents the results of (very) large-scale computational
tests, both analytical and numerical.
Section~5 presents a brief bestiary of MZVs and Euler sums, obtained from
Feynman diagrams that played vital roles in the inference of the conjectured
enumeration. A conclusion appears, with several open questions, in Section~6.

\subsection*{2. Conjectured enumerations}

It is conjectured that the answers to the three
questions, above, are
generated by
\begin{eqnarray}
\prod_{n\ge3}\prod_{k\ge1}(1-x^n y^k)^{E_{n,k}}&\stackrel{?}{=}&
1-\frac{x^3y}{1-x^2}\,\frac{1}{1-x y}\,,\label{A1}\\
\prod_{n\ge3}\prod_{k\ge1}(1-x^n y^k)^{M_{n,k}}&\stackrel{?}{=}&
1-\frac{x^3y}{1-x^2}\,,\label{A2}\\
\prod_{n\ge3}\prod_{k\ge1}(1-x^n y^k)^{D_{n,k}}&\stackrel{?}{=}&
1-\frac{x^3y}{1-x^2}+\frac{x^{12}y^2(1-y^2)}{(1-x^4)(1-x^6)}\,.\label{A3}
\end{eqnarray}

The solution to~(\ref{A1}) is that given by the author in~\cite{EUL},
in terms of what was there dubbed ``Euler's triangle'', with
the symmetric entries:
\begin{equation}
T(a,b)=\frac{1}{a+b}\sum_{d|a,b}\mu(d)\,P(a/d,b/d)\,,
\label{ET}
\end{equation}
where $P(a,b)=(a+b)!/(a!b!)$ are the entries in Pascal's triangle,
the sum is over all positive integers $d$ that divide both $a$
and $b$, and a M\"obius transformation is effected by
\begin{equation}
\mu(d)=\left\{\begin{array}{ll}
1&\mbox{ when $d=1$}\\
0&\mbox{ when $d$ is divisible by the square of a prime}\\
(-1)^k&\mbox{ when $d$ is the product of $k$ distinct primes}
\end{array}\right.\label{mu}
\end{equation}
which is the M\"obius function.
When $n$ and $k$ have
the same parity, and $n>k$, one obtains~\cite{EUL}
\begin{equation}
E_{n,k}=T(\mbox{$\frac{n-k}{2}$},k)\,.\label{A1s}
\end{equation}
With the exception of $\ln2$ and $\pi^2$ (which act as seeds)
all elements of the basis are thereby conjecturally enumerated.

The solution to~(\ref{A2}) is that given by BK. It too involves
Euler's triangle.
When $n$ and $k$ have
the same parity, and $n>3k$, one obtains~\cite{BK15}
\begin{equation}
M_{n,k}=T(\mbox{$\frac{n-3k}{2}$},k)\,.\label{A2s}
\end{equation}
With the exception of $\pi^2$ and $\zeta(3)$ (which act as seeds)
all elements of the basis are thereby conjecturally enumerated.

Conjecture~(\ref{A3}) appears in the final~\cite{BK15} version of BK,
though no solution to it was given there.
Here we motivate the Ansatz, and develop
the generators for irreducibles of specific depths, up to $k=6$.
Observe, first, that the order $y$ terms in~(\ref{A1})
merely assert the irreducibility of the 2-braid torus
knot-numbers~\cite{DK1}, $\zeta(2n+1)$ for $n>0$.
The claim of~\cite{EUL}, there supported
by a year of effort and $10^3$ CPUhours of testing,
is that all else is generated by including the factor of
$1/(1-x y)$ on the RHS, which gives~(\ref{A1s}),
by M\"obius inversion.
It then took little time for BK to discover a RHS in~(\ref{A2}) that was
consistent with all known data. To celebrate the absence of $\ln2$ from
the reduction of both MZVs and counterterms,
we omitted the factor of $1/(1-x y)$, obtained~(\ref{A2}),
and performed CPU-intensive tests on the resulting
prediction~(\ref{A2s}), which emerged unscathed.
It is important to remember the question thus answered:
how many Euler sums in the basis for MZVs?
It is ironical that the most complex of the three questions admits
of the simplest answer.
Consider weight-12 depth-2 irreducibles, with $M_{12,2}=2$.
The two basis elements may be taken as $\zeta(9,3)$ and
$\zeta(\overline{9},\overline{3})$. The latter is not an MZV,
but should be included, since it was shown in~\cite{EUL}
that $\zeta(4,4,2,2)$ reduces to a combination of
$\zeta(\overline9,\overline3)$ with MZVs of depths $k<4$.
Moreover, at weight 15 and depth 5, we found that $\zeta(6,3,2,2,2)$
reduces to a combination of $14\zeta(9,\overline3,\overline3)
-3\zeta(7,\overline5,\overline3)$ with MZVs of depths $k<5$.
We call this phenomenon ``pushdown''. The simplicity
of~(\ref{A2}) relies on its existence and regularity.

The final step, to the most complex of the three Ans\"atze,
in~(\ref{A3}), was taken more falteringly\footnote{A preliminary
version of BK got it wrong, by failing to model~(\ref{A3}) on the product
forms~(\ref{A1},\ref{A2}).}.
The question - how many MZVs in the basis for MZVs? -- is the most natural
to ask, in knot/number/field theory~\cite{DK}. One now
suspects that its answer will be the most complex of the three,
since one is denied the use of non-MZV terms, like
$\zeta(\overline{9},\overline{3})$, in the reduction of MZVs.
If pushdown makes~(\ref{A2}) so simple, then
its exclusion is expected to complicate~(\ref{A3}).
The final recipe of BK was the simplest possible, in these curious
circumstances: to add a term that encapsulates what was already
known about pushdown from depth 4 to depth 2, starting at weight 12.
The additional term was constructed, with frank naivety,
as follows. First, BK knew that the terms of order $y^2$ were given by
$\frac{x^{12}y^2}{(1-x^4)(1-x^6)}$, merely by supposing the correctness of
the conjectured~\cite{DZ,BBG} enumeration of irreducible depth-2 MZVs.
The factor of $(1-y^2)$ was then inserted on the grounds that pushdown
to depth $k=2$ originated from depth $k=4$.
This pushdown demonstrably~\cite{EUL,BGK} occurs at weight 12, where the
`rule of three', $D_{2n,2}=\lfloor\frac{n-1}{3}\rfloor$,
for irreducible depth-2 MZVs, in~\cite{BBG}, first differs from the
`rule of two', $M_{2n,2}=\lfloor\frac{n-2}{2}\rfloor$,
for 3-braid knots, in~\cite{BGK}.
It is supposed that pushdown to depth $k=2$ occurs
for greater even weights, at a rate that is metered~\cite{BGK}
by comparing~\cite{EUL} with~\cite{BBG}.

It should be stressed that recipe~(\ref{A3}) is the
simplest\footnote{The aforementioned false start was more complicated,
as it was not based on a product generator.} one that BK could concoct,
on the basis of depth-2 results.
Its genesis does not guarantee any success at
depths $k>2$.
It is hoped that the reader will share some the author's amazement at
the formidable success, below, of the highly specific
predictions at depths $k=3,4,5$.
To extract these predictions:
take the logarithm of each side of~(\ref{A3});
set $a=x^2$ and $b=x^3y$; expand in powers of $b$;
determine the generator for $D_{3k+2n,k}$ by M\"{o}bius inversion.
The result is
\begin{equation}
D_k(a)\equiv\sum_{n\ge0}D_{3k+2n,k}a^n
=\sum_{d|k}\frac{\mu(d)}{d}L_{k/d}(a^d)\,,
\label{dk}
\end{equation}
where
\begin{equation}
\sum_{k\ge1}L_k(a)b^k=\sum_{N\ge1}\frac{1}{N}\left(\frac{b}{1-a}
+\frac{b^2(b^2-a^3)}{(1-a^2)(1-a^3)}\right)^N.\label{lk}
\end{equation}
It follows that $D_k(a)$ is a ratio of polynomials, and that its
singularities occur exclusively at roots of unity. In particular:
\begin{eqnarray}
D_1(a)&=&\frac{1}{1-a}\,,\label{d1}\\
D_2(a)&=&\frac{a}{(1-a)(1-a^3)}\,,\label{d2}\\
D_3(a)&=&\frac{a(1 + a - a^2)}{(1-a)(1-a^2)(1-a^3)}\,,\label{d3}\\
D_4(a)&=&\frac{1+2a^2+a^3+a^4+2a^5+a^7-a^8}
         {(1-a)(1-a^3)(1-a^4)(1-a^6)}\,,\label{d4}\\
D_5(a)&=&\frac{1+2a+3a^2+3a^3+2a^4}
         {(1-a^2)^2(1-a^3)^2(1-a^5)}\,.\label{d5}
\end{eqnarray}
The complexity of the rational generators rapidly increases. For example,
\begin{equation}
D_6(a)=\frac{N_6(a)}{(1-a)(1-a^2)(1-a^3)(1-a^4)(1-a^6)(1-a^9)}\label{d6}
\end{equation}
has a numerator
\begin{equation}
N_6(a)=1+2a+3a^2+4a^3+6a^4+6a^5+6a^6+7a^7+4a^8+5a^9+4a^{10}+2a^{11}+2a^{12}
-a^{16}+a^{17}.\label{n6}
\end{equation}
As $n\to\infty$, a splendid behaviour of $D_{3k+2n,k}$ follows
directly from~(\ref{A3}):
\begin{equation}
k!\lim_{n\to\infty}n^{1-k}D_{3k+2n,k}=k\lim_{a\to1}(1-a)^k
D_k(a)=(3-\sqrt3)^{-k}+(3+\sqrt{3})^{-k}\,,\label{dinf}
\end{equation}
requiring in~(\ref{d6}) that $N_6(1)=52$, which
mental arithmetic shows to agree with~(\ref{n6}).

Already a spectacular success emerges: the generator~(\ref{d3})
is precisely that conjectured for depth-3 MZVs in~\cite{DZ,BG},
though {\em no\/} term of order $y^3$ occurs in the Ansatz~(\ref{A3}).
Thus the simple-minded input of~(\ref{d1},\ref{d2}) predicts~(\ref{d3}).
The obvious question remains: how does the conjecture fare
at greater depths? For $k>3$, data prior to BK was extremely scanty.
Personal communication from Don Zagier, reporting computations
performed by him and by Dror Bar-Natan, were consistent
with the first 5 terms predicted by~(\ref{d4}) at depth 4.
Warmed by this success,
we here extend it, by {\em very\/} large-scale computation,
to the first 16 terms. The method involves shuffle
identities~\cite{DZ,CK,LM,BBB}, outlined in the next section,
which provide rigorous upper bounds on irreducibles
at specific depths and weights.
Confirming that $D_{42,4}\le111$ involved 1.7 petashuffles.
Having thus tuned and tested the code,
we pushed on, to even larger systems of identities, at depth 5,
where the first 7 terms of~(\ref{d5}) have now been confirmed.
The challenge of testing a significant number of terms in~(\ref{n6})
was judged imprudent to attempt, unaided. Fortunately help is
at hand, from Jon Borwein, David Bradley and
Roland Girgensohn, in a collaboration~\cite{BBBR} hosted
by the Center for Experimental and Constructive
Mathematics (CECM) at Simon Fraser University, which is a node of
the Canadian High-Performance Computing Network.

\subsection*{3. Depth-length and weight-length shuffles}

Depth-length shuffles were used in~\cite{EUL,BBG,BG,BBB}. The idea
is simple, in the extreme, though less trivial to notate.
For simplicity, consider the product of a depth-1 Euler sum,
whose argument string is merely $\{s_1\}$, and an Euler sum of depth
$k-1$, with argument string $\{s_2,\ldots,s_k\}$. In the case of
alternating Euler sums, signs may be included, as bars~\cite{BK15,BBB}
or minus signs~\cite{EUL}, in these strings. Then the product
clearly entails the $k$ Euler sums~\cite{EUL} in which
$s_1$ is inserted, in all possible ways, in the other string.
The depth-length strings
\[
\{s_1,s_2,\ldots,s_k\}\,,\quad
\{s_2,s_1,\ldots,s_k\}\,,\quad\ldots\,,\quad
\{s_2,\ldots,s_1,s_k\}\,,\quad
\{s_2,\ldots,s_k,s_1\}\,,\quad\]
result, each with unit coefficient, together with sums of lesser
depth, which do not concern the present analysis of reducibility.
Generalization~\cite{BBBR} to the product of a string of depth $r$
with one of depth $k-r$ is as might be expected: take all
${k\choose r}$ shuffles
of the two depth-length strings that preserve the order of each.
Combined with the reducibility~\cite{BBB} of
\[\zeta(\mbox{argument string of length $k$})
+(-1)^k\zeta(\mbox{reversed string})\]
such depth-length shuffles appear to exhaust that
which can be concluded merely from the existence of a nested $k$-fold
sum; all else must take account of the form of the summand.

In addition to such depth-length shuffle identities (called permutation
identities in~\cite{EUL,BBG,BG}) there are weight-length shuffle identities
(corresponding to the partial-fraction identities of~\cite{EUL,BBG,BG}).
These follow from the existence of iterated-integral~\cite{CK} representations
for Euler sums. In the case of an MZV of weight $n$, a representation
has been given~\cite{DZ,CK,LM} in terms of an $n$-fold iterated integral
of the one-forms $d x/x$ and $d x/(1-x)$. The extension to alternating
sums was given in~\cite{BBB}: one has merely to include the one-form
$d x/(1+x)$, which adds a third character to the weight-length alphabet.

Weight-length shuffles are then expressions of the
ring structure of alternating~\cite{BBB} Euler sums, and their
restriction~\cite{DZ} to MZVs:
\begin{equation}
\zeta(\mbox{string}_1)\zeta(\mbox{string}_2)=\sum_{\mbox{shuffles}}
\zeta(\mbox{weight-length shuffled string})\,,
\label{shuf}
\end{equation}
where each shuffled string preserves the order of each of the two
constituent weight-length strings, in its iterated-integral representation.
The product of strings of weights $n_1$ and $n_2$ entails
${n_1+n_2\choose n_1}$ weight-length shuffles, each of which has a weight and
a depth that is the sum of those in the product, and many of which may occur
many times.

\subsection*{4. Computational testing}

For double sums,~(\ref{d2}) was the input to~(\ref{A3}).
However the correctness
of that input is still,
according to the author's understanding of~\cite{DZ,BBG}, a conjecture.
Thus it was tested, analytically, using REDUCE~\cite{RED},
to weight 44, which confirmed that the conjectured enumeration satisfied
the identities of~\cite{BBG}, and required no other.

For triple sums,~(\ref{d3}) was a spectacular output of~(\ref{A3}).
However the correctness of that output is still,
according to the author's understanding of~\cite{DZ,BG}, a conjecture.
Thus it was tested, analytically, also using REDUCE,
to weight 37, which confirmed that the conjectured enumeration satisfied
the identities of~\cite{BG}, and required no other.

The analytical results so obtained  were consistent with
the completeness and minimality of the concrete bases proposed
in~\cite{BBB}. Minimality is more or less
unprovable, since nothing, in principle, forbids further reductions,
beyond those entailed by shuffles. However, the lattice algorithm
PSLQ~\cite{DHB} resolutely failed to find anything new, wherever we
had the CPUtime and patience to probe.
More such tests may now be made with PSLQ,
since its author, David Bailey, has very recently
implemented Richard Crandall's
fast algorithm~\cite{REC} for computing Euler sums,
including alternations of sign, and achieving
a precision of 3200 digits at depth 5, and beyond.
Confirmation has been obtained of instances of conjectures made in~\cite{BBB},
and in the course of the present work. Results may shortly
appear in the high-performance computing literature. To date,
none of the many numerical discoveries, made with PSLQ, erodes
any conjecture in~\cite{BK15,EUL,BBB}.

The strategy adopted to test~(\ref{A3}) at depths 4 and 5,
with high computational efficiency, was to pre-process all depth-length
shuffles algebraically, using REDUCE, whose output was then converted to
FORTRAN code.
This was translated, by David Bailey's TRANSMP~\cite{DHB} code,
into multiple precision calls of MPFUN~\cite{DHB} routines,
which solved all the weight-length shuffles, modulo terms of lesser depth,
enabling reducibility analysis at a numerical precision which
provided overwhelming evidence that all known relations,
and no others, were satisfied.
The reader is (mercifully) spared programming details, which merely translate
the transparent idea, of exhausting all shuffle identities,
into computational practice. S/he is assured that the probability of
misidentification of numerical solutions of the identities
was less -- often many orders of magnitude less -- than $10^{-10}$.
The results that follow are hence not rigorously proven, though
we discount the slender possibility that they fail to deliver
the reductions that would have been achieved by computer
algebra, with integer arithmetic,
had one many gigabytes of core memory, and CPUyears of processing time,
to expend on the huge integers that would be generated.

A measure of what was involved is provided by the following
statistics. At depth $k=4$ and weight $n=42$, the ring structure~(\ref{shuf})
generates $1720620718074180$ shuffles (1.7 petashuffles)
each of which contributes unity to an element of a $16000\times10660$
matrix of constraints, whose rank deficiency was found to be 111.
At depth $k=5$ and weight $n=27$, the rank deficiency
of a $29900\times14950$ matrix
was found to be 36.

In total, we confirmed the emboldened rank deficiencies
in the integer sequence
\begin{equation}
{\bf 1,1,3,5,7,11,16,20,27,35,43,54,66,78,94,111},128,150,173,
196,224,254,284,\ldots\label{is4}
\end{equation}
generated by~(\ref{d4}), and those in the sequence
\begin{equation}
{\bf 1,2,5,9,15,23,36},50,71,96,127,165,213,266,333,409,498,600,720,851,
1005,\ldots\label{is5}
\end{equation}
generated by~(\ref{d5}). Neil Sloane has designated~(\ref{is4},\ref{is5})
as A019449 and A019450 in the on-line~\cite{NJAS} encyclopedia of integer
sequences.

For reasons that may be appreciated, the CPUtime required for the above
was considerable: about 20 CPUdays on a 256MB AlphaStation 600 5/333.
For that reason, testing of~(\ref{A3}) at depth 6 awaits input from
colleagues at CECM~\cite{BBBR}. Nor is this an idle pursuit, since at depth 6
there appear to be new~\cite{BBBR} features to solving
the depth- and weight-length shuffle identities.
If~(\ref{A3}) is not correct,
then depth 6 is the place to expect a failure. If a non-trivial
number of terms in the amazing numerator~(\ref{n6}) is, as hoped,
confirmed, then the chances of failure at depths $k\ge7$ appear remote, given
the essentially depth-2 nature of the input.

\subsection*{5. Euler sums: a field theorist's sample bestiary}

Lest Euler sums and MZVs appear a diversion
from serious-minded field theory, we briefly recall some of their
appearances in pQFT~\cite{BK15,EUL,DK2,BKP,BGK,5LB,1440,Z2S6,IMU3,LR,AFMT},
often in advance of their systematic investigation by the maths community.
To this traffic, from field theory to number theory, Dirk Kreimer has added
an equally noble trade route: from field theory to knot theory~\cite{DK}.

{\bf Depth-1}\quad Every pQFT practitioner knows that the odd zetas,
$\zeta(2n+1)$, are everywhere dense in multiloop QCD and QED.
In counterterms, they
correspond to 2-braid torus knots~\cite{DK1,DK2,DK3}. Here one beast is
gloriously absent~\cite{BK15} from pQFT: no subdivergence-free
four-dimensional counterterm can spawn $\pi^2$, for which there is no knot.

{\bf Depth-2}\quad A decade ago~\cite{1440},
before MZVs attracted apparent interest in the maths community,
the author failed to reduce a depth-2 weight-8 Euler sum, in the
$\varepsilon$-expansion of dressed two-loop diagrams.
The persistence~\cite{Z2S6}
of this state of affairs led to the prediction that it would surface
in 6-loop counterterms. It does~\cite{BKP}. It is the first
depth-2 irreducible MZV, enumerated by $D_{8,2}=1$.
Even earlier~\cite{Imu}, the St Petersburg group were unable to reduce
$\varepsilon$-expansions of critical exponents to depth-1.
It is now known why:
an infinite series~\cite{BGK} of depth-2 irreducibles
occurs, starting at weight 8. The corresponding knots start at
8 crossings, with the 3-braid~\cite{VJ} torus knot
$8_{19}=(4,3)$~\cite{BKP}.
Very recently~\cite{IMU3}, Anatoly Kotikov and the author have shown
that {\em alternating\/} Euler sums result from the $\varepsilon$-expansion
of critical exponents, in $D=3-2\varepsilon$ dimensions.
The first such beast is
$\zeta(\overline3,\overline1)$, which occurs with a multiple of
$\zeta(4)$ that is process dependent. Thus odd-dimensional counterterms
mimic even-dimensional massive diagrams, for it was observed in~\cite{EUL}
that two of the most important 3-loop results in pQFT -
for the electron anomaly~\cite{LR} and the $\rho$-parameter~\cite{AFMT} -
have precisely that structure in their weight-4 terms.

{\bf Depth-3}\quad Likewise, there
was an obstacle to the reduction of triple sums of weight 11,
obtained by cutting vacuum diagrams~\cite{5LB}.
The irreducible occurs in the 7-loop
beta-function of $\phi^4$-theory~\cite{BKP}. As at depth 2~\cite{BBG},
mathematicians later encountered~\cite{DZ,BG} this phenomenon.
The associated knot is the sole positive 11-crossing 4-braid~\cite{BKP}.
Its uniqueness is required by $D_{11,3}=1$, and is confirmed by
the knot enumeration of BK.

{\bf Depth-4}\quad Perhaps the most far-reaching gift from pQFT to number
theory occurred at depth 4 and weight 12~\cite{EUL,BGK}. Despite untiring
effort, BK kept encountering an apparent mismatch between
field, knot and (the then current state of) number theory.
The (indubitably correct) result of~\cite{BBG}, that $D_{12,2}\le1$,
jarred with the existence of a {\em pair\/}~\cite{EUL}
of 12-crossing 3-braid knots in the counterterms~\cite{BGK} of pQFT.
The resolution was dramatic: the second knot corresponds to $\zeta(4,4,2,2)$,
which is pushed down~\cite{EUL} to $\zeta(\overline9,\overline3)$,
in the simpler enumeration~(\ref{A2}),
giving $M_{12,2}=D_{12,2}+D_{12,4}=1+1=2$.
Without this stimulus from pQFT, the author
would not have discovered~\cite{EUL} such pushdowns, and the simplicity
of~(\ref{A2}) might have remained hidden.
Instead of seeking a simple answer to the (now seen to be) difficult
question {\bf Q3}, BK backtracked to {\bf Q2}, which has the simplest answer
of the three. Then it was possible to return to {\bf Q3}, input a minimal
amount of information, interpreted in the light of Richard
Feynman~\cite{BK15,BGK} and Vaughan Jones~\cite{VJ},
observe the success~\cite{DZ,BG} of~(\ref{A3}) in giving~(\ref{d3}),
and arrive, via~(\ref{d4},\ref{d5}), at the predicted
sequences of deficiencies in~(\ref{is4},\ref{is5}), tested here,
non-trivially, at a cost of several petashuffles.

Such are the fruits of computational field theory.

\newpage\subsection*{5. Conclusion and questions}

The conclusion is easy to state: the conjectured
enumeration~(\ref{A3}) emerged in the course of calculations in
field theory~\cite{BK15,EUL,BGK}, is illuminated by
knot theory~\cite{DK1,DK2,DK},
and is shown by~\cite{BBG,BG,BBBR}, and by the emboldened results
in~(\ref{is4},\ref{is5}),
to be in great shape,
at all depths $k<6$, however striking that may appear
from the point of view of number theory~\cite{DZ,BBB,MEH}.

Further questions are easier to pose
than to answer:
\begin{enumerate}
\item[{\bf Q4}]\quad Why?
\item[{\bf Q5}]\quad Is~(\ref{n6}) correct, at depth 6?
\item[{\bf Q6}]\quad Is there an alternative to the CPU-intensive testing
procedures, adopted here?
\end{enumerate}
These, and related issues, are receiving due
attention~\cite{4TR,BK4,IMU3}.

\subsection*{Acknowledgements}
Of the many colleagues whose generosity of spirit has aided this work,
I single out for especial thanks: David Bailey, of NASA-Ames,
whose PSLQ~\cite{DHB} algorithm made computational number theory
a joy; Jon Borwein, of CECM, for welcoming me to experimental mathematics;
Bob Delbourgo, for a safe physics haven in Hobart, where this work began;
Tony Hearn, of RAND, whose REDUCE~\cite{RED}
made computational knot theory much easier than I had supposed;
Chris Stoddart, at the Open University, for management of a petashuffle-proof
AlphaCluster; Don Zagier, for guiding my rash intrusion into his field with
true grace, and for suggesting that any guess be cast in product form.
Combined with these, David Bradley, Richard Crandall, Roland Girgensohn,
John Gracey, Tolya Kotikov and Neil Sloane
formed a valued network of advisors. At heart, I relied on the intuition,
patience and tenacity of Dirk Kreimer, which provided daily sustenance.

\raggedright

\end{document}